\documentstyle[12pt,prb,aps]{revtex}

\input epsf

\begin{document}
\draft
\title{Pairing evidence of 8CB molecules adsorbed on MoS$_{2}$:
Influence of 2D commensurability on the intralamellar structure }
\vskip6mm
\author{E. Lacaze$^{1}$,  M. Alba$^{2}$, M.
Goldmann$^{3,4}$, J.P. Michel$^{1}$ and F. Rieutord$^{5}$}
\address{$^{1}$ Groupe de Physique des Solides, Universit\'es Paris 7
et 6,
 UMR-CNRS 75-88, 2 place Jussieu, 75251 Paris CEDEX, France
 \vskip1mm
 $^{2}$ DRECAM/SPEC, bat 772, CE-Saclay, 91191 Gif sur Yvette CEDEX,
France
 \vskip1mm
 $^{3}$ LURE,  Bat 209D, Universit\'e Paris Sud, 91405  Orsay CEDEX ,
France
 \vskip1mm
 $^{4}$ GRPB,Universit\'e Paris 5, 45 rue des Saints P\`eres, Paris
CEDEX 6, France
 \vskip1mm
 $^{5}$ CEA-Grenoble, DRFMC/SI3M/MCI, 17 rue des Martyrs, 38054
 Grenoble
 CEDEX 9, France}

\maketitle

\vskip8mm
\begin{abstract}
By combining X-ray diffraction studies under grazing incidence (GIXD)
and Scanning Tunneling Microscopy (STM) measurements, we have
precisely determined the structure of 8CB molecules adsorbed
on MoS$_{2}$, under the thick organic film. The
commensurability of the adsorbed network and the intracell
structure have been determined, revealing a complex structure
characterized by a non-equivalence of adjacent lamellae with pair
associations of molecules in one lamella over two. We have interpreted such
a result by a
simple model of a single lamella. This pair
association in one lamella over two appears as a direct
evidence of the connection between the commensurabilities in the two
directions. The
value of the molecule-substrate potential corrugations is
particularly high, indicating that the dipolar momentum of 8CB
molecules could play a fundamental role in the molecule-substrate
interactions.\\
\end{abstract}
\pacs{PACS 61.10.-i,68.35.Bs, 61.30, 68.35.Md}

\section{Introduction}
Since the last decade, various interfaces between organic films and
crystals have been described as well ordered in 2D, whereas
the organic bulk is obviously liquid or liquid crystal\cite{Mar00}. This is
even
true on chemically inert crystals, such as graphite or MoS$_{2}$ where
molecules are only physisorbed, as probed by a number of Scanning
Tunneling Microscopy (STM) measurements
\cite{Fos88,Har90,Rab93,Wal96,Taki99}. These interfaces are
interesting in particular for two reasons. First, they can be studied
very precisely at ambient temperature, as presented in the present article,
which has
led to the possibility of connecting the microscopic structure of
the first adsorbed molecules to the relevant surface interactions;
second, the structure is studied under the organic film which can
lead in a second step to studies of the propagation of  interfacial
ordering into the organic
bulk. In case of liquid crystal phases (8CB bulk is smectic at
ambient
temperature), this phenomenum is called anchoring \cite{Jer91}.
In the case of physisorbed rare gas atoms or
simple molecules on the same substrates,
isotropic structures are formed, so that
equilibrium
structures can be exactly calculated \cite{Vil83}.
They are
described as commensurate or
incommensurate with respect to the substrate.  These
structures can be generally described with one or two molecules in the
crystallographic cell, such that the structure is completely
determined by its commensurability. Organic molecules are more
complex and exact calculations are already extremely more
difficult to perform. However their study in the adsorbed configuration can be
considered
as a first step towards
even more complex adsorbed molecules as for example biological
molecules.

On graphite or  MoS$_{2}$ substrates, organic molecules are usually
organized in
lamellae which can be straight or
regularly kinked \cite{Iwa90,Lac97}.
Until now, the molecule-substrate interactions have been considered
in some
numerical calculations \cite{Cle94} or through a phenomenological
Landau de
Gennes-type model \cite{Lac97}. This mean-field model describes the
monolayer structure as lamellae, which periodicity is imposed by
the substrate, ignoring the intra-lamella structure. Structures
are described as commensurate (straight lamellae) or
incommensurate (kinked lamellae) with respect to the substrate,
depending
mainly on the ratio between the substrate-molecule and
molecule-molecule interactions and on the mismatch between the
natural
periodicity of the lamellae and the one imposed by the substrate. This
model allows a global interpretation of the STM
results and leads to a
  commensurate  description of  various structures which
are naively expected to be incommensurate on such substrates since the
molecules-substrate interactions are expected to be weak, in particular
weaker than
on metallic substrates \cite{Hosh94,For94,Sou98}. In order to
confirm such
results, we have studied a straight lamellae interface, which is ``a
priori'' commensurate,
as probed by STM \cite{Har90}: 4-8-alkyl-4'-cyanobiphenyl (8CB) molecules
adsorbed
on molybdenum disulphide (MoS$_{2}$) substrate. STM
images of 8CB on MoS$_{2}$ \cite{Har90,Lac97} reveal that the
molecules
are flat on the
substrate, organized in straight lamellae, with a head to tail
geometry. By combining two techniques,
STM and X-ray
diffraction at grazing incidence (GIXD), we have verified the
commensurability of the interfacial structure and we
have also precisely determined the intralamellar
structure, allowing to go beyond the phenomenological
model and to propose a study of the
relationship
between microscopic structures and interface interactions in
physisorbed molecules.

The combination of
STM and GIXD experiments appeared particularly useful for the
determination of such adsorbed structures. Single monolayers only
are often studied in case of adsorbed organic molecules
\cite{Sou98,Wal98,Mey99}.
They can be observed by
combining LEED and Auger experiments. In the case of an interface
between a substrate and a thick organic film, these LEED and AUGER
measurements are
unoperating due to the presence of the bulk and mainly
two techniques remain available, STM and GIXD. However it is difficult to
image
simultaneously
the substrate and the
adsorbed molecules by STM, especially in case of a semi-conducting
substrate as
MoS$_{2}$ (by contrast to graphite as substrate \cite{Wal98}). Another
limitation of STM techniques is due to drift
phenomena of the
piezoelectric materials used for scanning the sample
surface, preventing a precise estimation of
absolute distance values.
Finally, as the contrast in STM images is not perfectly
understood,
the precise molecular structure is difficult to infer
from the images. On the other hand the Bragg peaks coming from the
monolayer and from the substrate structures can be simultaneously obtained
from GIXD
experiments. This
allows a precise
determination of the relative
orientation of the
two structures as well as of the lattice constants and of the molecule
localizations
 which can be extracted with a precision as good as 0.5
$\AA$.
The major disadvantage of this
latter technique remains the complexity of the experiment mainly
due to the small amount of matter contained in  physisorbed organic
monolayers which imposes the use of
synchrotron sources. This complexity justifies that up to now mainly STM
experiments
have been
performed on such interfaces. However the refinement of X-ray
data can be
strongly simplified if one introduces constraints on the molecule position
parameters. These constraints can be obtained from STM
images,
leading then to a fruitful combination of the two techniques. To our
knowledge
the only physisorbed organic film
studied by combined STM and GIXD has been the 10CB/graphite system, showing
metastable commensurate structure,
coexisting with
the a priori more stable incommensurate one\cite{Dai93}.
In the first part of this article, we describe the experimental
details; in a second
part, the X-Ray diffraction results; in a third part, we analyse the
peak intensities trough a model of the molecules localization in the
monolayer crystallographic cell. Finally in a fourth part, we
discuss the energetic balance between molecule-molecule and molecule
substrate interactions  by developing an unilamellar model of the
monolayer structure.
\section{Experimental}
The 8CB film has been
prepared by melting the organic material on top of a freshly
cleaved MoS$_{2}$ substrate, at
80$^{o}$C. The 8CB is a BDH product used without any purification.
MoS$_{2}$ natural single crystal comes
from Australia (Queensland) through the Wards' company. This lamellar
compound can be easily cleaved, giving a surface parallel to the basal
planes. The surface plane is composed of sulphur atoms organized with
an  hexagonal symmetry (3.16$\AA$ as cell parameter),
with
less than 0.02$^{o}$ mosaicity, as checked by X-Ray
diffraction. Grazing incidence X-Ray diffraction experiments
have been performed on beam-lines BM32 at ESRF and D41 at LURE. The
incident
beam (8 keV) reaches the interface through the 8CB
bulk at an incidence of 0.5$^{o}$, comparable to the MoS$_{2}$
critical angle and larger than the 8CB bulk one. The diffracted
intensity
is scanned parallel to the plane by a linear detector covering an
angle of 13$^{o}$ (PSD) perpendicular to the
surface at
LURE-D41 and a
solid state
detector at ESRF-BM32. The in-plane resolution is of the order of
1 mrad. The beam size width is
about 500 $\mu$m and its intensity is monitored by a diode. The STM and GIXD
experiments have been performed on many different samples. Always
reproducible results have been obtained, but only the most
intense Bragg
peaks of the monolayer structure were observed at LURE-D41.
\section{Results}
31 in-plane diffraction peaks (Table I, fig. 1) of the adsorbed molecules
have been
measured. They can be indexed in a (4x8) superstructure
position, indicating that the
monolayer structure is indeed commensurate with the MoS$_{2}$
lattice.
However such a unit cell contains only one lamella, and a close
inspection of the STM
images indicates that the number of non equivalent lamellae is 4. Considering
this constraint, the
basic unit corresponds to a (4x32) superstructure and each
peak has been indexed in this
(4x32) superstructure (fig. 1). The low index Bragg peaks of such a
superstructure
can not be observed since they are
very close to the direct beam.
An example of diffraction peak, (4,-48), is presented on the figure
2. The rocking curve (fig. 2a) indicates that the in-plane mosaicity
(0.02$^{o}$) mimics
the substrate one, showing that the adsorbed monolayer
is constituted of well-ordered single crystallized domains.
The observed radial scan width (fig.2b) is limited by the
experimental resolution, revealing a correlation
length at least
larger than 1000 $\AA$.

The experimental intensities measured on a given sample (BM32
experiment) are indicated on Table 1. They have been obtained by
a gaussian adjustment to the peak shapes.
The reciprocal space have been explored  along the twelve
 high
 symmetry directions and four main commensurate  positions  have been
 unambiguously assigned to a nul intensity after careful investigations. Since
the $\theta$-range is
rather small ($\Delta \theta$$\leq$13$^{o}$), we neglected
any geometrical
correction. We used these
intensity values to perform a refinement of the molecular structure:
we start with an an eight-molecule cell, built in order to
form lamellae parallel to the MoS$_{2}$ [100]
direction, with a head to tail geometry. The cell
parameters
are
fixed to
the (4x32) MoS$_{2}$ superstructure. Calculated intensities are
adjusted
to the measured ones with a least square criterium. Molecules are
considered as flat on
the substrate, therefore only 2D calculations are performed.
They are
divided in two rigid parts: the alkyl
chain in trans configuration and the cyanobiphenyl part also flat on the
substrate. The degrees of freedom of the model are the length of the alkyl
chain and of the cyanobiphenyl group (considered as equal for all
molecules), the angles between the alkyl
chain and the cyanobiphenyl group (with an equal value
for the molecules of same lamella), the
location and the overall rotation of  molecules. Considering the STM
observations, and in order to limit the
number of
parameters, we divide the crystallographic cell in
two blocks of four molecules, which can be translated one from each
other, the value of the displacement being an adjustable parameter.

We consider two other fit parameters, anisotropic
Debye-Waller terms and the
monolayer 3-fold degeneracy. Indeed, due to the hexagonal symmetry
of the MoS$_{2}$, three different directions (at $\pm$60$^{o}$) are
equivalent
for the adsorbed 2D crystal. They can coexist on the sample,
and their relative weight must be adjusted. Note that
since the commensurabilities along the two high symmetry
 directions of the  MoS$_{2}$ substrate are multiples from each other,
Bragg peaks originating from two coexisting domains can be
superimposed.

The best result ($\chi^{2}$ = 3.19) is presented on the figure 3a. The
corresponding calculated Bragg peaks intensities,
compared to the measured ones, are presented on the table I. The alkyl
chain length is 0.866 nm, the cyanobiphenyl
length 0.901 nm and the angles between them are 34.5$^{o}$ for
 one
lamella and 35.2$^{o}$ for
the other one. The displacement between the two four molecules blocks
has been obtained as half the cell parameter. The crystallographic cell
is then a "centered"
c(4x32) superstructure of the MoS$_{2}$. Considering that the 8CB
cell is eight times longer along the k direction than along the h
direction, an STM images can be normalized in order to avoid the
deformations connected to the drift of the piezoelectric materials and
superimposed to the
fit solution. Such a
comparison, presented on the figure 3b, shows that the
refinement
result is in very good agreement with the STM observations, in
particular the different orientations and locations of the eight
molecules in the cell.

\section{Discussion}
\subsection{Molecular conformation}
We have allowed independent values for the angles between the alkyl
chain and the cyanobiphenyl group of the
molecules of two
successive lamellae. Values very close from
each other are finally obtained, also similar to the
one calculated for an isolated flat molecule, 35.5$^{o}$.
This indicates that the adsorbed 8CB molecules on MoS$_{2}$ are
essentially
not distorted, which is coherent with
the numerical calculation of 8CB molecules adsorbed on graphite
\cite{Cle94}, but differs from the recent observations on
organic molecules covalently bonded onto Ag(111) \cite{My99}. The obtained
lengths of the alkyl chain and of the cyanobiphenyl group are close
(about 10$\%$
larger) to
the ones estimated for an isolated flat
molecule in a
completely trans conformation. This confirms that the
adsorbed
molecules are lying flat on the substrate, with the alkyl chain,
completely elongated, in all trans conformation. This indicates that,
in their isotropic phase, the molecules are
mobile enough to find their 2D-crystallized
equilibrium conformation, with only very few gauche defects \cite{Sou98}.
Due to the number of measured
Bragg
peaks, we could not consider the possibility of rotation of
the phenyl groups, but the good agreement between the measured and
calculated Bragg intensities indicates that, if such a rotation
occurs,
it remains small. The observed flat conformation of the molecules is
clearly related
to a strong molecule-substrate interaction.
\subsection{Unilamellar model}
The observation of a pair association in one lamella over two is probably
the most spectacular result of these
X-ray measurements (fig. 3). We underline that such a
non-equivalence of two successive lamellae, with the presence of
pair-associations in one lamella and equidistant molecules in the
neighboring one, is a particularly robust
result which appears almost systematically in the fit solutions.
Avoiding pair
association lead to very high $\chi^{2}$ values. It demonstrates that,
whatever organized in a commensurate
network, the adsorbed molecules do not lie on identical adsorption
sites. The origin of
such a feature is probably
connected to the
microscopic interactions between molecules as well as between
molecules and the substrate, which can not be described in a
mean field
model. In order to check this assumption, we have built a simple
model for a single lamella. We consider a commensurate model, with  two
dipoles
which can be displaced anywhere in the cell, with an
antiferroelectric alignment. The knowledge of the 8CB
characteristics
(dipole (D = 4.9 D) \cite{Gue86,Cle94}, polarizability
($\alpha$ = (40 10$^{-30}$ 4$\pi$$\epsilon$$_{o}$) m$^{3}$) \cite{Mit92},
ionization
potential (I = 8.7eV, derived from the diphenyl one by analogy with the
hydrocarbons case) \cite{Gut67}
allows the calculation of the energy per molecule as a function of the
distance,
r, between the molecule and its first neighbours \cite{Israel91}:
\begin{eqnarray}
    E  && = E_{D} + E_{VdW} + E_{S} \nonumber\\
       && = -{ D^{2} \over 4 \pi \epsilon_{o} }
      \left( {1 \over r^{3}} +  {1\over (4a-r)^{3}} \right) \nonumber\\
      && -{3 \alpha^{2} I\over \left (4 (4 \pi \epsilon_{o}\right) ^{2})}
      \left( {1 \over r^{6}} +  {1\over (4a-r)^{6}} \right) \nonumber\\
      && + C \left( {1 \over r^{12}} +  {1\over (4a-r)^{12}} \right)
\nonumber\\
      && + E_{o} - B \cos 2\pi\left({2a-r\over 2a }\right)
 \end{eqnarray}
with r in $\AA$ and the
energy in J.
E$_{D}$ is the term of dipolar interactions between adjacent molecules.
E$_{VdW}$ is the term of
Van der Waals interactions  between adjacent molecules which includes the
dispersion forces as
well as a steric repulsion where C is a phenomenological value as
usual. The attractive dispersion interactions appear
to be about three to four times larger than the attractive dipolar ones, by
varying the
distances between molecules from 4.3 to 6.32$\AA$, as already pointed
out by Tildesley
et al. \cite{Cle94} [note1]. E$_{S}$ is the molecule-substrate
potential which is described by a constant term E$_{o}$ for the adsorption
energy, and an oscillating term, the
potential corrugations, which period is fixed to the MoS$_{2}$ cell
parameter a = 3.16$\AA$.

Plotting the energy as a function of r leads to two types of
configuration, depending on the substrate corrugations parameter, B (fig.
4). If B is strong, one minimum minimorum appears, corresponding to a
location of the molecules in the substrate potential wells (fig. 4c). The
intermolecular
distance is then r$_{1}$ = 2a = 6.32$\AA$.
If B is weak, two r values give equal energy minima and pair
associations are formed (fig. 4a). The molecules are located out of the
potential wells and the intermolecular distance is mainly imposed
by the balance between attractive dipolar, dispersion interactions and steric
repulsion. One can then calculate C by neglecting B $\cos$ 2$\pi$(2a-r/2a):
we have determined through our X-ray refinement a pair intermolecular distance
of 4.3$\AA$ (r$_{2}$) which leads to C = 5.088 10$^{-12}$J
$\AA$$^{12}$.\\
A first order transition between
the two configurations occurs for a critical value B$_{c}$
(B$_{c}$ = 7.45 10$^{-20}$J) which can
be calculated by equating the two energy minima at r$_{1}$ and
r$_{2}$ (fig. 4b):
\begin{eqnarray}
    B_{c} && =
 {1\over\displaystyle 1-\cos 2\pi\left({ 2a-r_{2}\over\displaystyle 2a}
 \right)}\star \nonumber\\
    \Bigg[  &&-{ D^{2} \over 4 \pi \epsilon_{o} }
\left( {2 \over r_{1}^{3}} - {1 \over r_{2}^{3}} -  {1\over
(4a-r_{2})^{3}} \right) \nonumber\\
&& -{{3 \alpha^{2} I\over \left (4 (4 \pi \epsilon_{o}\right) ^{2})}}
\left( {2 \over r_{1}^{6}} - {1 \over r_{2}^{6}} -  {1\over
(4a-r_{2})^{6}} \right) \nonumber\\
&& + C \left( {2 \over r_{1}^{12}} - {1 \over r_{2}^{12}} -  {1\over
(4a-r_{2})^{12}}  \right) \Bigg]  \nonumber\\
 \end{eqnarray}
The observation of the two kinds of solutions on the same
sample, one lamella with pair associations, adjacent to one lamella
with equidistant molecules, indicates a variation of the
substrate corrugation parameter from B$\leq$B$_{c}$ to B$\geq$B$_{c}$
between two successive lamellae. This has to be
correlated to the refinement results which
indicates an average difference in molecular
orientations of around 10$^{o}$ for two neighboring lamellae. This
directly
demonstrates the anisotropic nature of the molecule-substrate
corrugations which depend on the orientation of the molecule with
respect to the substrate
crystallographic
directions. We can also conclude that the calculation of B$_{c}$ gives a
minimum
value for the
substrate corrugation which indeed stabilizes the observed configuration with
equidistant
molecules: 7.45 10$^{-20}$J per molecule  (i.e. 17 k$_{B}$T) . We
checked that this value is independent of the precise form of the
repulsion function and remains identical for different type of repulsive
functions (for example varying as
C/r$^{9}$ or C/r$^{15}$).
\subsection{commensurability of the lamellae}
The explanation for the observed alternate solution needs to go beyond
the unilamellar model described above.
It can be justified by considering the 2D structure
associated to a single type of lamella. If one considers a c(4x32)
superstructure composed by
equidistant molecules only, a limited number of possibilities exists due
to
the commensurability of the molecules location. It appears then that,
in each case, the
cyanobiphenyl groups of adjacent lamellae would be located
directly in front of each other. Such a c(4x32) superstructure
suffers then
from strong steric incompatibilities
and a c(4x34) superstructure appears to be the
smallest
superstructure which does not exhibit such a steric repulsion.
However in a c(4x34) superstructure the mismatch between the
lamellae period imposed by the substrate and the natural lamellae
period would jump from 2.5\% to 9\%. Such a lack
of compacity would
become unfavorable for the attractive part of the
interactions between adjacent lamellae: indeed 11CB molecules on MoS$_{2}$
with a 10\% similar mismatch relax by creating kinks leading to an
incommensurate structure perpendicularly to the lamellae
\cite{Lac97}. Alternating two intralamellar structures
favours the attractive interactions without inducing steric
repulsion
due to  different locations of  the cyanobiphenyl groups of
adjacent lamellae. On the other hand, a 2D structure composed of
pair associations
only would correspond to a completely incommensurate structure within
the
lamellae. Consequently, in our model the postulated commensurability is
in fact imposed by the interaction between
adjacent lamellae. The observed alternating structure
appears as a compromise which allows to preserve the
commensurability in both directions, within the lamellae and between
the lamellae.
The energy gain of the perpendicular commensurability is balanced
by the presence of pair
associations, a priori less favorable, one
lamella over two. It shows how the
commensurabilities of these
structures in both directions are strongly correlated and cannot be
described through a
phenomenological model which ignores the intralamellar structures. It
justifies to perform now similar
X-ray
studies on 9CB and 11CB on MoS$_{2}$, since our result suggests that 9CB
and 11CB lamellae on MoS$_{2}$ which are
kinked \cite{Iwa91,Lac97}, should present a single type of lamellae
with a commensurate intra-lamellar
structure.
\subsection{Molecule-substrate interaction}
The obtained value of the substrate corrugations is high, 7.45 10$^{-20}$J (17
k$_{B}$T) and such, consistent
with the observation of a commensurate structure at ambiant
temperature and to
its thermal stability [note2]. Very few
data exist for evaluating it more precisely, especially in case of
MoS$_{2}$. Best of available
studies
deal
with rare gas atoms on graphite. In case of xenon atoms on
graphite, the substrate corrugation has
been estimated at about 2.78 10$^{-21}$J \cite{Ste73,Deb86}, 27 times smaller
than
our estimated value of 8CB on MoS$_{2}$.

Substrate-atoms interactions are only of the  Van der Waals
type
in the Xe/graphite system. Calculating the homogeneous Van der Waals part
of the
molecule-substrate interactions through a continuous
model \cite{Israel74}, one obtains
a value 15 times
smaller for the Xe/graphite system \cite{Taf65,Erg68} than for the
8CB/MoS$_{2}$ one \cite{Eva65,Lia71} instead of the
27 times for the substrate potential corrugations ((0.25
$\alpha$$_{Xe}$)/D$^{3}$J compare to (0.38
$\alpha$$_{8CB}$)/D$^{3}$J, with D the distance in $\AA$
between
the Xe or
the 8CB and the substrate and $\alpha$ the polarizability)
\cite{Mit92,Israel91}. However
the 8CB/MoS$_{2}$-Xe/graphite corrugations ratio is expected to be
smaller than the 8CB/MoS$_{2}$-Xe/graphite homogeneous
potentials ratio, due to the large and anisotropic size of the 8CB which
should average and such diminish the substrate corrugations compare to
the Xe atoms.
In the way
to explain the calculated increasing,
one can assume that the previous lamellar model
overestimate the corrugation parameter, B. Calculating
the
dipolar interaction for two 8CB molecules, as they are located in the
structure, one obtains only a 1.06\% increase with respect to the
model result. However, Van der Waals
interactions between the molecules may have been overestimated due to
the far location of the more
polarizable cyanobiphenyl groups in the antiferroelectric alignment.
Another possibility is that the 8CB/MoS$_{2}$ corrugation potential
is not only related to
pure Van der Waals interactions. This is supported by the presence of a
strong dipolar momentum, 4.9D,
located on the
cyanobiphenyl group in 8CB molecules possibly
inducing special molecule-substrate
interactions, apart the pure Van der Waals ones. Dipole-induced
dipole
interactions have a negligible influence compare to Van der Waals
ones, as shown by continuous calculations (8.10$^{-20}$/D$^{3}$ J
compare to 2.43 10$^{-18}$/D$^{3}$ J). However one should consider that
MoS$_{2}$ is
composed of two
types of
atoms, which differences in polarizabilities could play a
role. Numerical calculations could test such an hypothesis. Clearly such an
effect should be absent on graphite, as
shown in the study of polar
molecules as CH$_{2}$Cl$_{2}$ adsorbed on graphite \cite{bah94}.
This result on MoS$_{2}$ suggests that the
location of the cyanobiphenyl groups with respect to this substrate
should
play a predominant role as it
is generally assumed that the alkyl chains determine the structure on
the graphite substrate.

\section{Conclusion}
We have precisely determined, through a combination of scanning
tunneling
microscopy and
grazing incidence X-rays diffraction, the 2D network of 8CB molecules
adsorbed on
MoS$_{2}$ which anchor along the [100] MoS$_{2}$ direction. The
crystallographic cell has been determined, as well as the intracell
fine structure. The X-ray diffraction spectra
refinement lead a molecular structure very
close to the STM images.
Introduction of constraints in the refinement method allows a reduction of
the a priori
large number of fit parameters, connected to the size of the
molecules
and to the number of molecules in the unit cell.
The constraints
being directly
defined by the STM results, this demonstrates how the combination
between the two techniques (STM and GIXD) is fruitful.
In the reverse way, the X-ray results evidence the fine structures
within the lamellae, which are not straightforward in the STM
images.

The structure appears to be commensurate, perpendicularly to the
lamellae as predicted by a phenomenological
model \cite{Lac97}, but moreover within the lamellae.
The 8CB
crystallographic cell
is a centered
c(4x32) MoS$_{2}$ superstructure

The determination of the crystallographic parameters is clearly not
sufficient to
describe correctly the
adsorbed network. The description of the intracell structure
evidences complex intralamellar structures that we have interpreted through
an unilamellar model of
the microscopic interactions. Pair
associations in one lamella over
two correspond to lamellae with
molecules out of the molecule-substrate potential wells adjacent to
lamellae with
molecules inside the wells. This alternating series reflect the connection
between
the commensurabilities along and perpendicularly to the lamellae.
The critical value of the molecule-substrate potential
corrugations which impose the observed commensurate lamellae has
been calculated to be 7.45 10$^{-20}$J per molecule. This particularly high
value
with respect to the Xe/graphite system could
be connected to the presence of a high dipolar momentum in the 8CB
molecules. The anisotropy of these corrugations is already
demonstrated, directly connected to the orientation of the molecules
with
respect to the substrate crystallographic directions. Numerical
simulations as well
as similar
measurements on
slightly different systems could now be very usefully compared to such
a study, in order to go towards the complete understanding of
the complex interactions substrate-organic molecules.
This could allow, in the future, of directly forecasting adsorbed structures
of molecules, at least in case of physisorption.

\appendix
$\it{Note 1}$:

This observation justifies the
occurrence of two different structures for only slightly different
molecules on
the same substrate: a "single layer" one for 8CB on MoS$_{2}$, which
lamellae are characterized by an antiferroelectric alignment, a
priori
favorable for dipolar interactions; a "double layer" one for 10CB on
MoS$_{2}$ \cite{Iwa92},  which
lamellae are characterized by a ferroelectric alignment,
unfavorable
for dipolar interactions \cite{Taki99}. The relative lost of dipolar
energy
is even smaller for larger molecules (10CB compare to 8CB) and could
be partly
compensated by a win in Van der Waals interaction connected to the
proximity
of the more polarizable cyanobiphenyl
groups in the second
structure. So the win in
molecule-substrate
interactions necessary for imposing a "double layer" structure can be
only small and indeed
correlated to some odd-even effect due to the position of the end of
the alkyl chain on the underlying substrate, as recently proposed
\cite{Taki99}.
This remark also justifies the observation that in 10CB/graphite case
the two types
of structure have been evidenced, also through the association of STM
and GIXD experiments \cite{Dai93}.

\appendix
$\it{Note 2}$:

The 8CB adsorbed layer Bragg peaks starts to disappear
at 120$^{o}$C only, whereas the 8CB bulk already starts to evaporate
around 100$^{o}$C.

\vskip10mm
\narrowtext
\begin{table}
\begin{center}
\begin{tabular}{||l||c|r||}
\hline
Bragg peak&I$_{m}$/I$_{o}$&I$_{c}$/I$_{o}$\\
\hline
 (0,8,0) & 0.16 & 0.12 \\
 (1,-8,0) & 0.2 & 0.2 \\
 (0,14,0) & 0 & 0 \\
 (2,-10,0) & 0 & 0 \\
 (2,0,0) & 1 & 1 \\
 (2,-16,0) & 0.69 & 0.83 \\
 (0,-16,0) & 0.81 & 0.85 \\
 (0,16,0) & 0.81 & 0.85 \\
 (2,-24,0) & 0.03 & 0.05 \\
 (3,-16,0) & 0.02 & 0.12 \\
 (2,16,0) & 0.06 & 0.05 \\
 (4,-16,0) & 0.04 & 0.13 \\
 (3,8,0]) & 0.04 & 0.04 \\
 (3,-32,0) & 0.09 & 0.07 \\
 (4,-24,0) & 0.03 & 0.07 \\
 (3,-40,0) & 0.03 & 0.05 \\
 (5,0,0) & 0.01 & 0.01 \\
 (2,32,0) & 0.20 & 0.14 \\
 (4,-48,0) & 0.31 & 0.28 \\
 (4,16,0) & 0.42 & 0.41 \\
 (6,-32,0) & 0.49 & 0.61 \\
 (6,-16,0) & 0.28 & 0.29 \\
 (0,48,0) & 0.08 & 0.09 \\
 (6,0,0) & 0.01 & 0 \\
 (6,-48,0) & 0.60 & 0.60 \\
 (-7,0,0]) & 0 & 0 \\
 (2,48,0) & 0.04 & 0.04 \\
 (6,16,0) & 0 & 0 \\
 (8,-16,0) & 0.03 & 0.03 \\
 (5,40,0) & 0.02 & 0.02 \\
 (5,48,0) & 0.02 & 0.02 \\
\end{tabular}
\vskip20mm
\caption{8CB Bragg peaks intensities normalized to the (2,0,0) peak:
I$_{m}$/I$_{o}$ measured
intensities, I$_{c}$/I$_{o}$
calculated
intensities.}
\label{table1}
\end{center}
\end{table}

\begin{figure}
\vskip50mm
\begin{center}
\epsfxsize=16cm
\epsffile{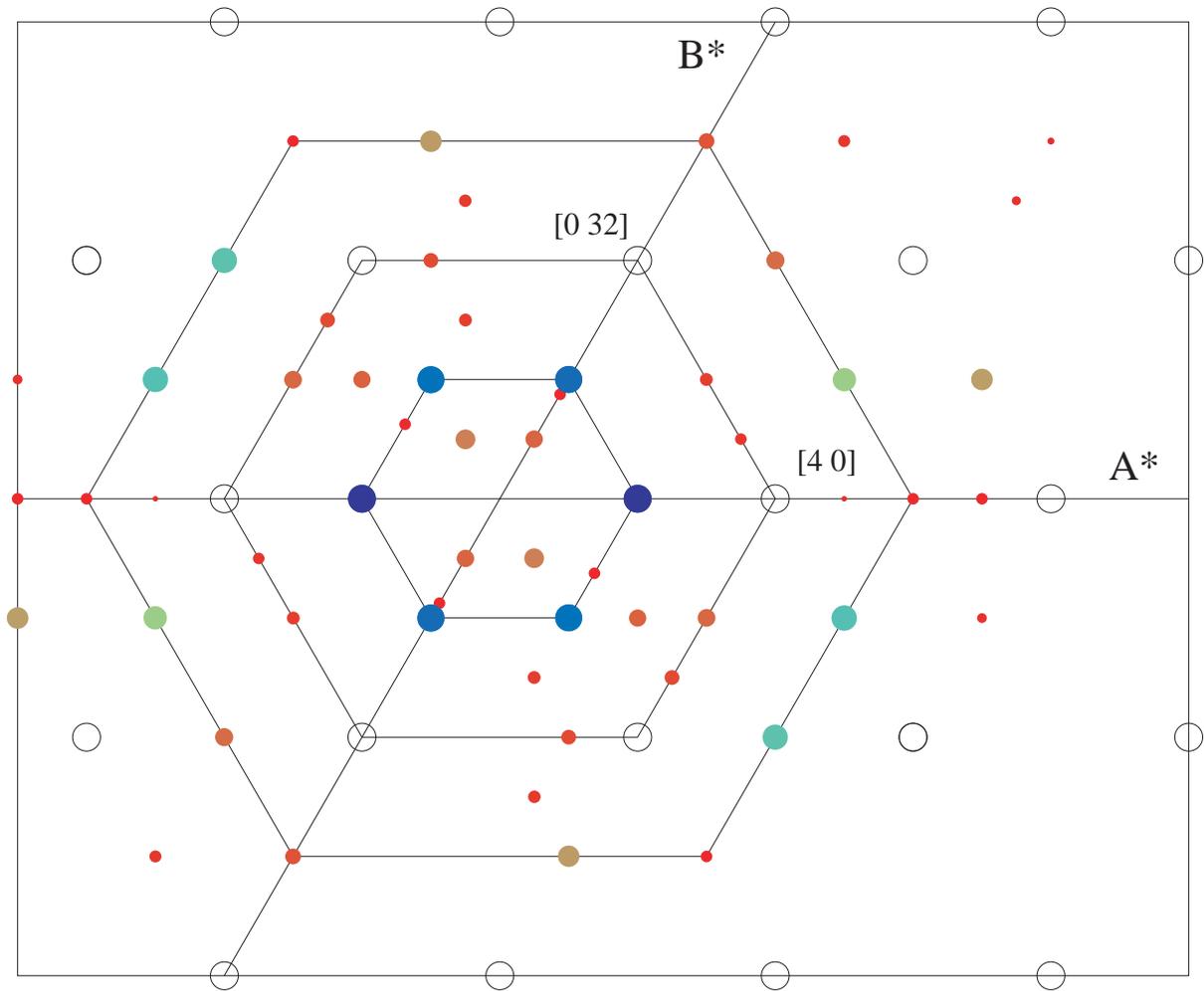}
\vskip20mm
\caption{Reciprocal space map of the 8CB/MoS$_{2}$ network. Open
circles correspond to the MoS$_{2}$ bragg peaks and full ones to the
8CB ones which are indexed in the (4x32) superstructure of MoS$_{2}$. The
intensities are qualitatively represented by the size of the
circles.}
\label{figure1}
\end{center}
\end{figure}

\begin{figure}
\begin{center}
\epsfxsize=14cm
\epsffile{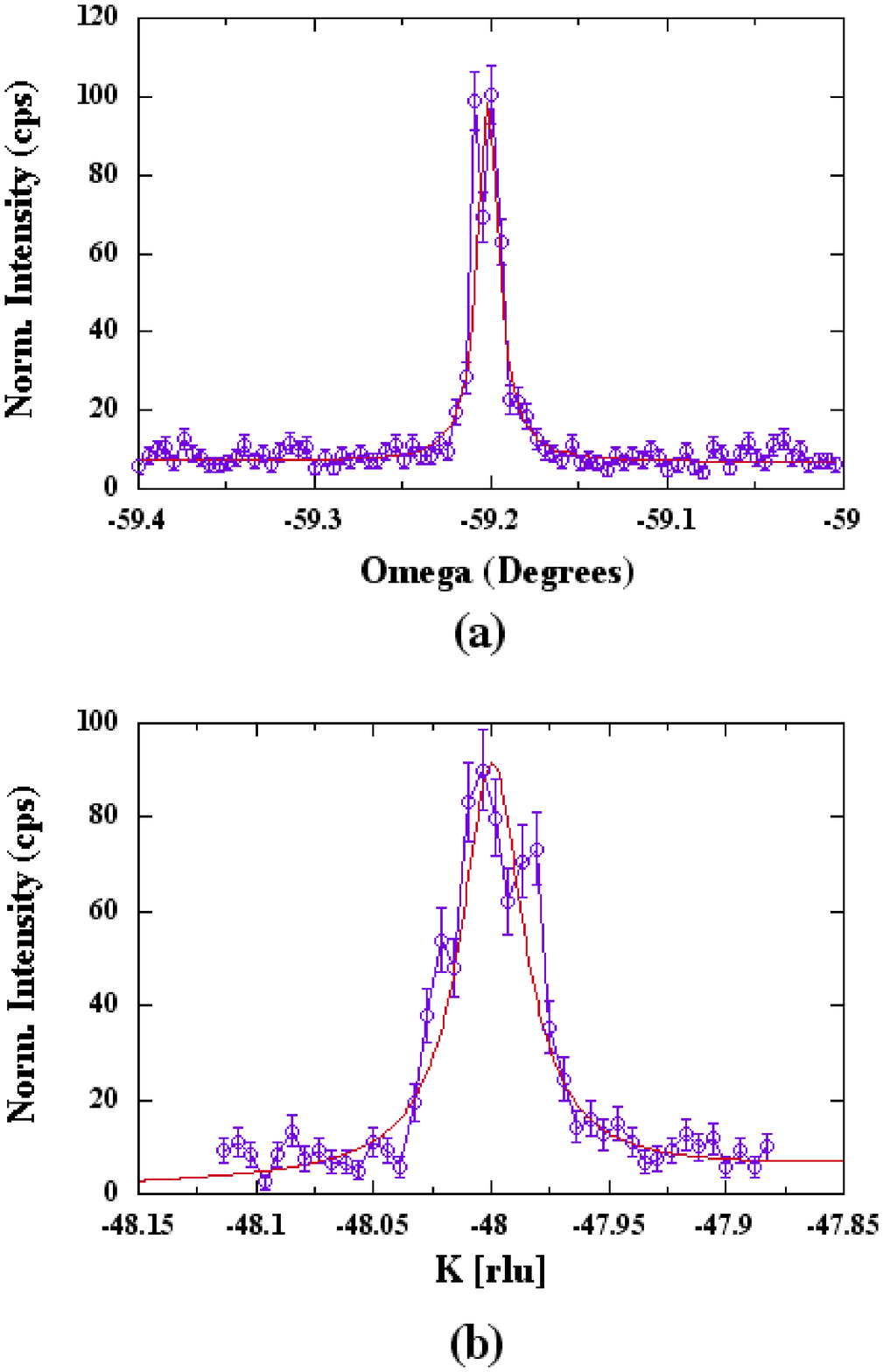}
\vskip4mm
\caption{ a) rocking curve of the (4,-48,0) bragg peaks which
indicates
a mosaicity of 0.02$^o$ (the intensity, in count per second, is normalized
to the incident beam
intensity).
b) radial curve of the (4,-48,0) bragg peaks which corresponds
to a correlation length at least larger than 1000$\AA$.}
\label{figure2}
\end{center}
\end{figure}

\begin{figure}
\begin{center}
\epsfxsize=6cm
\epsffile{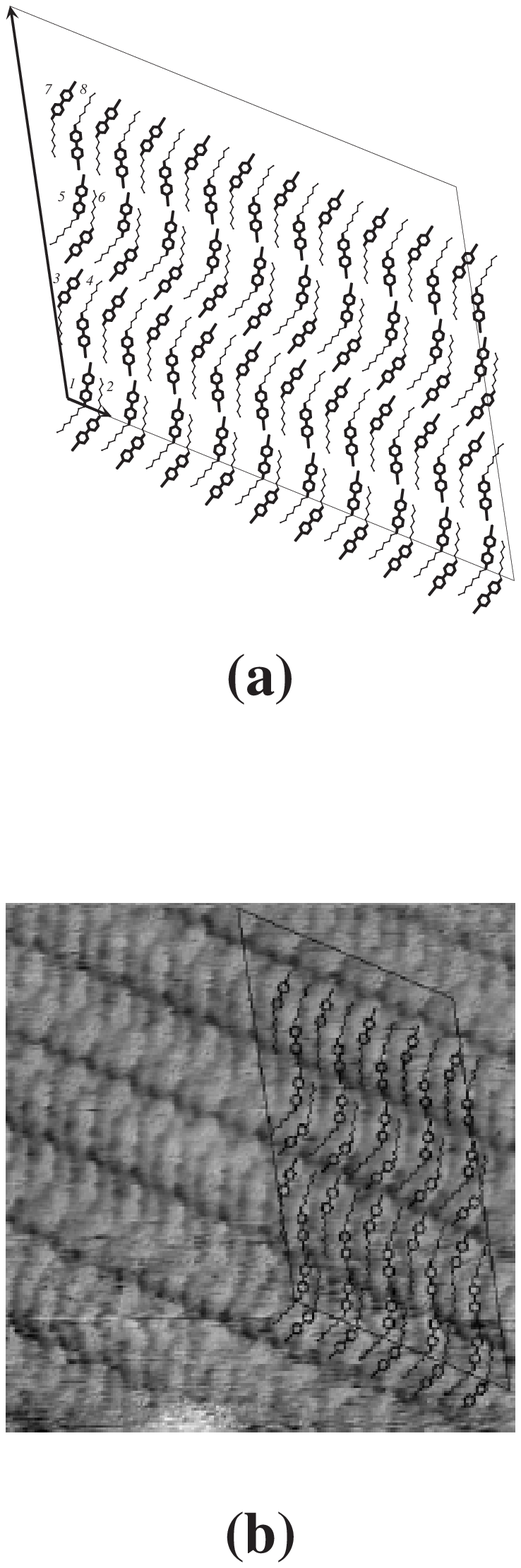}
\vskip6mm
\caption{ a) result of the refinement of the X-ray data with the
eight molecules of the crystallographic cell (the
weights of the three possible orientations are 0.42 along the [100],
0.21 along the [010] and 0.37 along the [1-10]; the Debye-Waller
coefficients are 0
along the [100] and 0.02 along the [010]). Pair associations are formed
between the molecules 1/2 and 5/6.
b) comparison between the refinement and the STM image, rescaled.}
\label{figure1b}
\end{center}
\end{figure}

\begin{figure}
\begin{center}
\epsfxsize=12cm
\epsffile{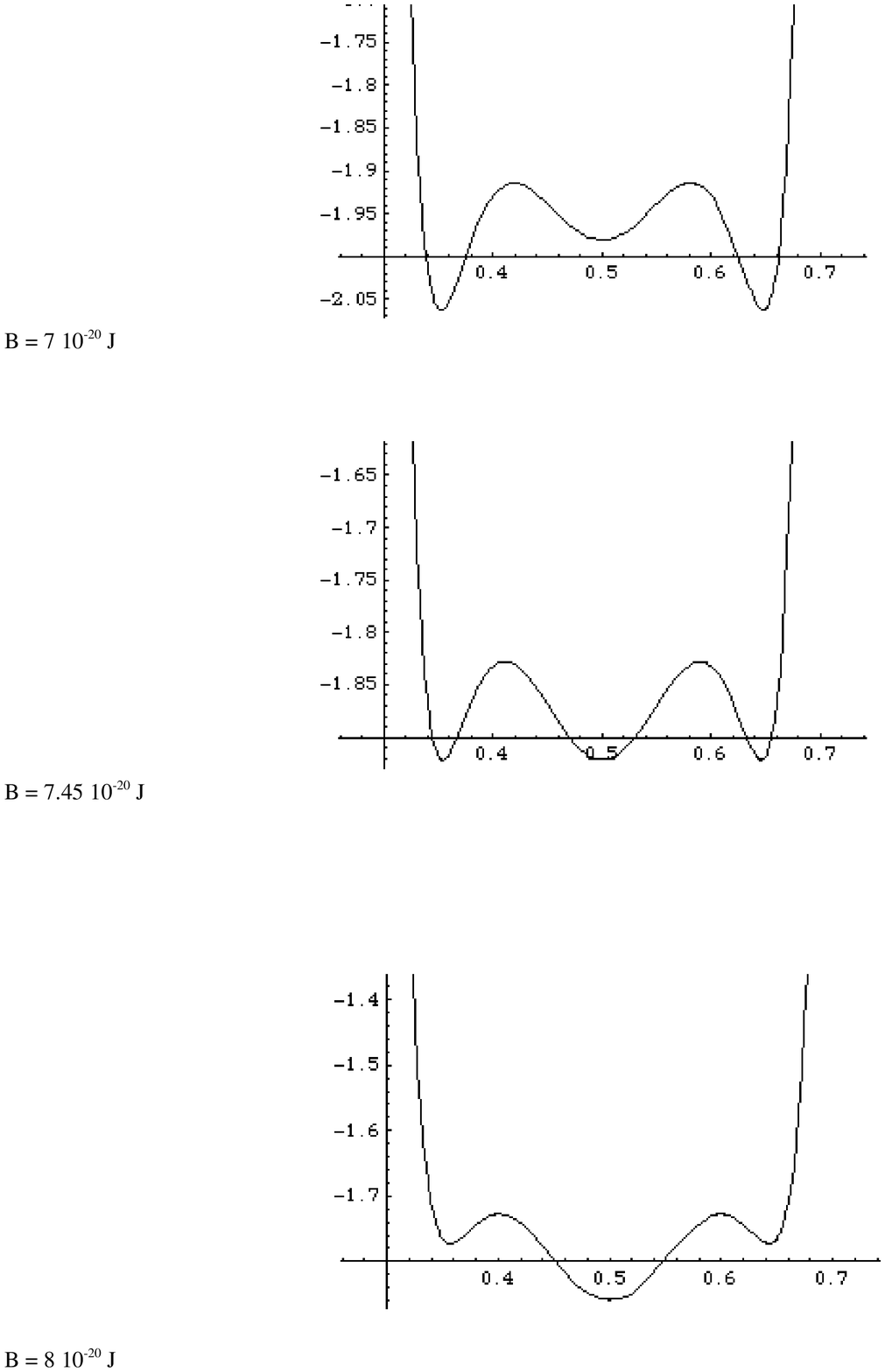}
\vskip12mm
\caption{energy (over B) of an 8CB molecule adsorbed on MoS$_{2}$ surface,
versus r ($\AA$)
for various values of the corrugation potential, B: B = 7
10$^{-20}$J, 7.45 10$^{-20}$J and 8 10$^{-20}$J.}
\label{figure2a}
\end{center}
\end{figure}

\end{document}